# Bloch-type magnetic skyrmions in two-dimensional lattice

Wenhui Du, Kaiying Dou, Ying Dai*, Baibiao Huang, Yandong Ma*

School of Physics, State Key Laboratory of Crystal Materials, Shandong University, Shandanan Street 27, Jinan 250100, China

*Corresponding author: daiy60@sina.com (Y.D.); yandong.ma@sdu.edu.cn (Y.M.)

**Abstract**

Magnetic skyrmions in two-dimensional lattice are a prominent topic of condensed matter physics and material science. Current research efforts in this field are exclusively constrained to Néel-type and antiskyrmion, while Bloch-type magnetic skyrmions are rarely explored. Here, we report the discovery of Bloch-type magnetic skyrmions in two-dimensional lattice of $MnInP_2Te_6$, using first-principles calculations and Monte-Carlo simulations. Arising from the joint effect of broken inversion symmetry and strong spin-orbit coupling, monolayer $MnInP_2Te_6$ presents large Dzyaloshinskii-Moriya interaction. This, along with ferromagnetic exchange interaction and out-of-plane magnetic anisotropy, gives rise to skyrmion physics in monolayer $MnInP_2Te_6$, without needing magnetic field. Remarkably, different from all previous works on two-dimensional lattice, the resultant magnetic skyrmions feature Bloch-type, which is protected by $D_3$ symmetry. Furthermore, the Bloch-type magnetic bimerons are also identified in monolayer $MnTlP_2Te_6$. The phase diagrams of these Bloch-type topological magnetisms under magnetic field, temperature and strain are mapped out. Our results greatly enrich the research on magnetic skyrmions in two-dimensional lattice.

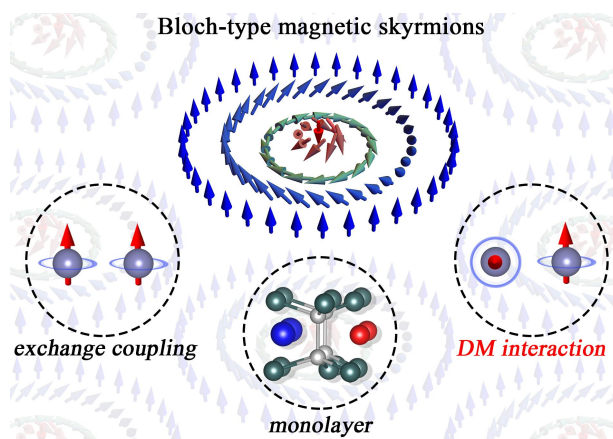

**Introduction**

Magnetic skyrmions are nanoscale chiral spin textures with nontrivial real-space topology [1-5]. The topological nature is accompanied with an integer topological charge Q and decoupled from the crystal lattice. Since the initial experimental observation in 2009 [6], this swirling configuration has attracted great attention and spurred rapid development due to a variety of exotic characteristics, such as topologically robust against defects, solitonic feature and efficient current-driven dynamics [7,8]. These appealing properties provide not only new opportunities for exploring fundamental topological physics but also great potential for future spintronics devices [9-11]. In recent years, with the rise of two-dimensional (2D) materials [12-16] and discovery of magnetism in monolayers $CrI_3$ [17], $VSe_2$ [18,19] and $Fe_3GeTe_2$ [20], many efforts are devoted to investigating skyrmion physics in 2D lattices, and magnetic skyrmions in 2D lattices have been the topic of intense research in condensed matter physics and material science [21-35].

According to the inner spin arrangements of vorticity ω and helicity γ, magnetic skyrmions can debut in three typical cases of (ω, γ) = (1, 0 or π), (1, ±π/2) and (-1, 0 or ±π/2 or π), corresponding to Néel-type, Bloch-type and antiskyrmion, respectively [36]. These distinct magnetic skyrmions are equivalently attractive and exhibit fertile physics. The variation of magnetic skyrmions generally arises from Dzyaloshinskii-Moriya interaction (DMI) [37,38]. Such short-range antisymmetric exchange interaction, in fact, acts as a chiral interaction and fixes the specific rotational sense of spins. Driven by various bulk and interface DMI, all the three types of magnetic skyrmions are extensively explored in conventional B20 bulk and thin-film systems [39-48]. Nevertheless, magnetic skyrmions reported in 2D lattice, in most cases, are constrained to Néel-type [21-31], and only a few 2D systems with antiskyrmion are also reported [32,33]. Actually, up to now, Bloch-type magnetic skyrmions in 2D lattice which requires special DMI have not been reported yet. From the perspective of fundamental research, it is essential to explore such Bloch-type magnetic skyrmions in two-dimensional lattice.

In this work, through first-principles calculations and Monte-Carlo (MC) simulations, we report the identification of Bloch-type magnetic skyrmions in 2D lattice of $MnInP_2Te_6$. Due to the combined effect of broken inversion symmetry and strong spin-orbit coupling (SOC), large DMI is obtained in monolayer $MnInP_2Te_6$. The competition between DMI and ferromagnetic exchange interaction as well as out-of-plane magnetic anisotropy results in skyrmion physics in $MnInP_2Te_6$ in the absence of magnetic field. Importantly, protected by $D_3$ symmetry, the resultant magnetic skyrmions feature the intriguing Bloch-type. Meanwhile, Bloch-type magnetic bimerons are also observed in monolayer MnTlP$_2$Te$_6$. In addition, the evaluations of these Bloch-type topological magnetisms under magnetic field, temperature and strain are systematically discussed.

**Results and Discussion**

**Figure 1(a)** presents the crystal structure of monolayer MnXP$_2$Te$_6$ (X=In, Tl), which is isostructural to MnPTe$_3$ [49]. Its unit cell contains one Mn, one X, two P and six Te atoms. Each Mn/X atom is octahedrally coordinated with six Te atoms, forming a hexagonal lattice with the space group P312 (D$_3$); see **Figure 1(b)**. Obviously, the inversion symmetry of monolayer MnXP$_2$Te$_6$ is broken. The lattice constants of monolayer MnXP$_2$Te$_6$ are summarized in **Table S1**. The dynamic and thermal stabilities of monolayer MnXP$_2$Te$_6$ are confirmed by performing the phonon spectra calculations (**Figure S1**) and AIMD simulations (**Figure S2**), respectively.

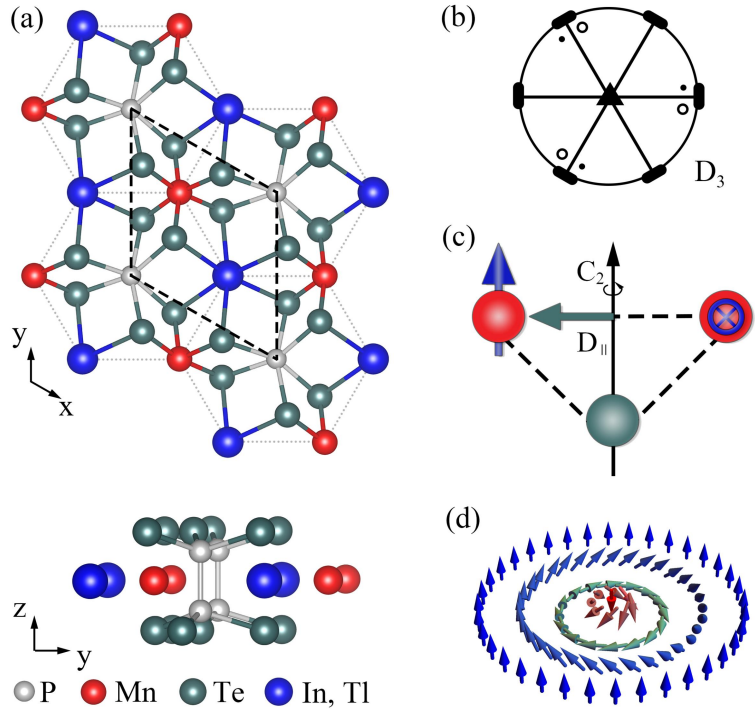

**Figure 1**. (a) Crystal structure of monolayer MnXP$_2$Te$_6$ from top and side views. (b) Equatorial plane projection of the point group for monolayer MnXP$_2$Te$_6$. (c) Schematic illustration of DMI acting on the adjacent magnetic moments of monolayer MnXP$_2$Te$_6$. (d) Schematic representation of Bloch-type spin configuration.

The valence electron configurations of Mn and In/Tl atoms are 3d$^5$4s$^2$ and 5s$^2$5p$^1$/6s$^2$6p$^1$, respectively. Upon coordinating with the neighboring Te atoms, each Mn atom donates three electrons, while each X atom denotes one electron, which would generate the formal oxidation states of +3 and +1 for Mn and X atoms, respectively. Note that under distorted octahedral coordination environment, Mn-3$d$ orbitals roughly split into two groups, i.e., two high-lying e$_g$ ($d_{xy}, d_{x^2-y^2}$) and three low-lying t$_{2g}$ ($d_{xz}, d_{yz}, d_{z^2}$) orbitals. As $d$ electrons of Mn favor high-spin configuration, according to the Hund's rule and Pauli exclusion principle, the four left electrons would half-fill three t$_{2g}$ and one e$_g$ orbitals. This would yield a magnetic moment of 4 µ$_B$ on each Mn atom. Our first-principles calculations indeed show that the magnetic moment per unit cell is 4 µ$_B$ and mainly localized on Mn atoms (see **Figure S3**).

To characterize the detailed interaction between the magnetic moments of monolayer MnXP$_2$Te$_6$, we introduce the following Heisenberg spin Hamiltonian:

$$H = \sum_{\langle i,j \rangle} \bm{D}_{ij} \cdot (\bm{S}_i \times \bm{S}_j) - J \sum_{\langle i,j \rangle} \bm{S}_i \cdot \bm{S}_j - \lambda \sum_{\langle i,j \rangle} S_i^z S_j^z - K \sum_i (S_i^z)^2 - mB_a \sum_i S_i^a \quad (1)$$

Here, $\bm{S}_i$ is unit vector representing local spin of Mn atom at site $i$. $\langle i,j \rangle$ represents nearest-neighboring sites. $\bm{D}_{ij}$, $J$, $\lambda$ and $K$ stand for parameters of DMI, Heisenberg isotropic exchange, anisotropic symmetric exchange and single ion anisotropy, respectively. The last term is the Zeeman energy, where $m$ and $B_a$ indicate on-site magnetic moment of Mn atom and external magnetic field along the $a$ ($a$ = x, z) direction, respectively. In view of the D$_3$ symmetry, as shown in **Figure 1(c)**, there is a two-fold rotation axis vertically passing through the middle of the bond between two neighboring Mn atoms. According to the Moriya's rule [38], the DMI vector $\bm{D}_{ij}$ for the nearest-neighboring Mn atoms is perpendicular to this two-fold rotation axis, which can be expressed as $\bm{D}_{ij} = d_\parallel \bm{u}_{ij} + d_\perp \bm{z}$ with $\bm{u}_{ij}$ representing the unit vector pointing from site $i$ to $j$ and $\bm{z}$ being the out-of-plane unit vector. Due to the D$_3$ symmetry, the out-of-plane component $d_\perp$ presents a staggered arrangement for the six nearest-neighboring Mn atoms, which renders it negligible in average. It is worth emphasizing that the in-plane component $d_\parallel$ is parallel to the bond between two adjacent Mn atoms. It makes the spins rotating in a plane perpendicular to the radial direction; see **Figure 1(d)**. This is different from most of the previous works wherein the DMI forces the spins rotating in a plane along a radial direction [21-31]. As we will show below, such particularly DMI favors Bloch-type magnetic skyrmions. To obtain the in-plane component $d_\parallel$, the left- and right-hand spin-spiral configurations are considered, as shown in **Figure 2(a)**. Strikingly, $d_\parallel$ in monolayer MnInP$_2$Te$_6$ (MnTlP$_2$Te$_6$) is calculated to be 2.87 (5.79) meV, which comparable with or even remarkably larger than that reported in previous studies [21-34].

To get more insights into the large DMI in monolayer MnXP$_2$Te$_6$, we investigate the atomic resolved SOC energy difference $\Delta E_{soc}$ between two different spin-spiral configurations. As shown in **Figure 2(c)**, the DMI is mainly contributed by Te atom, while other atoms make only slight contributions. This suggests that the Fert-Levy mechanism [50,51] plays a dominant role for inducing DMI in monolayer MnXP$_2$Te$_6$. By considering the strong SOC strength within Te atom, we can understand why large DMI is obtained in both systems. Another point we can see from **Figure 2(c)** is that the magnitude of $\Delta E_{soc}$ for Te atom in monolayer MnTlP$_2$Te$_6$ is pronouncedly larger than that in monolayer MnInP$_2$Te$_6$. This might be related to the relatively large atomic radius of Tl atom, which exacerbates the bipartite asymmetry in monolayer MnTlP$_2$Te$_6$ and thus leads to the large magnitude of $\Delta E_{soc}$ and DMI.

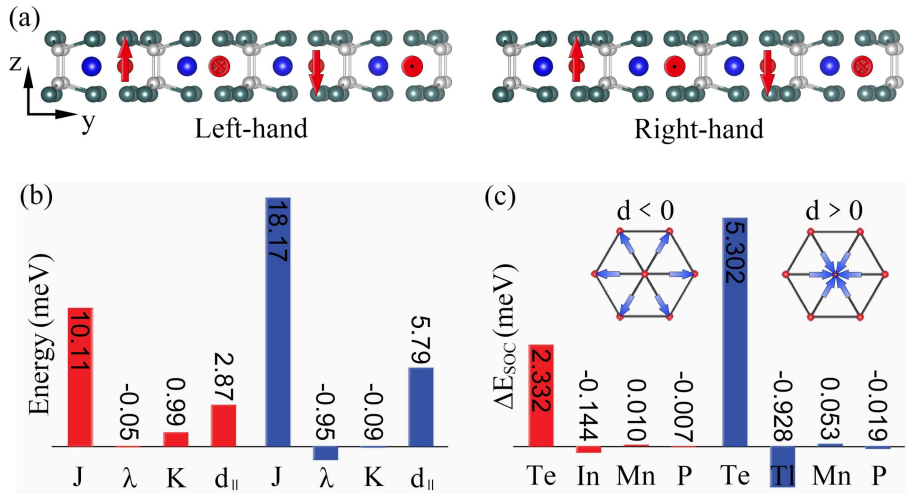

**Figure 2**. (a) Left and right-hand spin-spiral configurations used to obtain the in-plane DMI parameter $d_{\parallel}$ and atomic projected SOC energy. (b) The calculated $J$, $\lambda$, $K$ and $d_{\parallel}$ of monolayer MnXP$_2$Te$_6$. (c) SOC energy projected on atoms ($\Delta E_{soc}$) for monolayer MnXP$_2$Te$_6$. Insets in (c) show the in-plane DMI vectors $d_{\parallel}\boldsymbol{u}_{ij}$ (blue vectors) between the nearest-neighboring Mn atoms (red balls).

To obtain $J$, $\lambda$ and $K$, we considered four different magnetic configurations as displayed in **Figure S4**. The isotropic exchange parameter $J$ is calculated to be 10.11 and 18.17 meV, respectively, for monolayer MnInP$_2$Te$_6$ and MnTlP$_2$Te$_6$. This indicates FM exchange interaction is favorable for neighboring magnetic atoms for both systems. The single ion anisotropy $K$ consists of two parts: one is magnetocrystalline anisotropy $K_{MCA}$ and another is magnetic shape anisotropy $K_{MSA}$, i.e., $K = K_{MCA} + K_{MSA}$. The former arises from SOC, while the latter results from magnetostatic dipole-dipole interaction and usually prefers in-plane magnetization. For monolayer MnInP$_2$Te$_6$, $K$ and $\lambda$ are calculated to be 0.99 and -0.05 meV, respectively, which prefer out-of-plane and in-plane magnetizations. Note that magnetic anisotropy is determined by the combined effect of $K$ and $\lambda$. Obviously, the positive $K$ is much larger than $\lambda$, which indicates that the spins in monolayer MnInP$_2$Te$_6$ tend to align along the out-of-plane orientation. Different from monolayer MnInP$_2$Te$_6$, monolayer MnTlP$_2$Te$_6$ has $K = -0.09$ meV and $\lambda = -0.95$ meV, suggesting the preference of in-plane magnetization. Based on the obtained $J$, $\lambda$, and $K$, we study the Curie temperature $T_C$ of monolayer MnInP$_2$Te$_6$ and MnTlP$_2$Te$_6$, which are estimated to be 130 and 230 K, respectively (**Figure S5**). **Figure S6** shows the band structures of monolayer MnXP$_2$Te$_6$. It can be seen that the spin-up bands cross the Fermi level, while the spin-down bands lie away from the Fermi level. This leads to a fully spin polarization for the conduction electrons around the Fermi level. Therefore, monolayer MnXP$_2$Te$_6$ exhibits 2D ferromagnetic half-metallic nature.

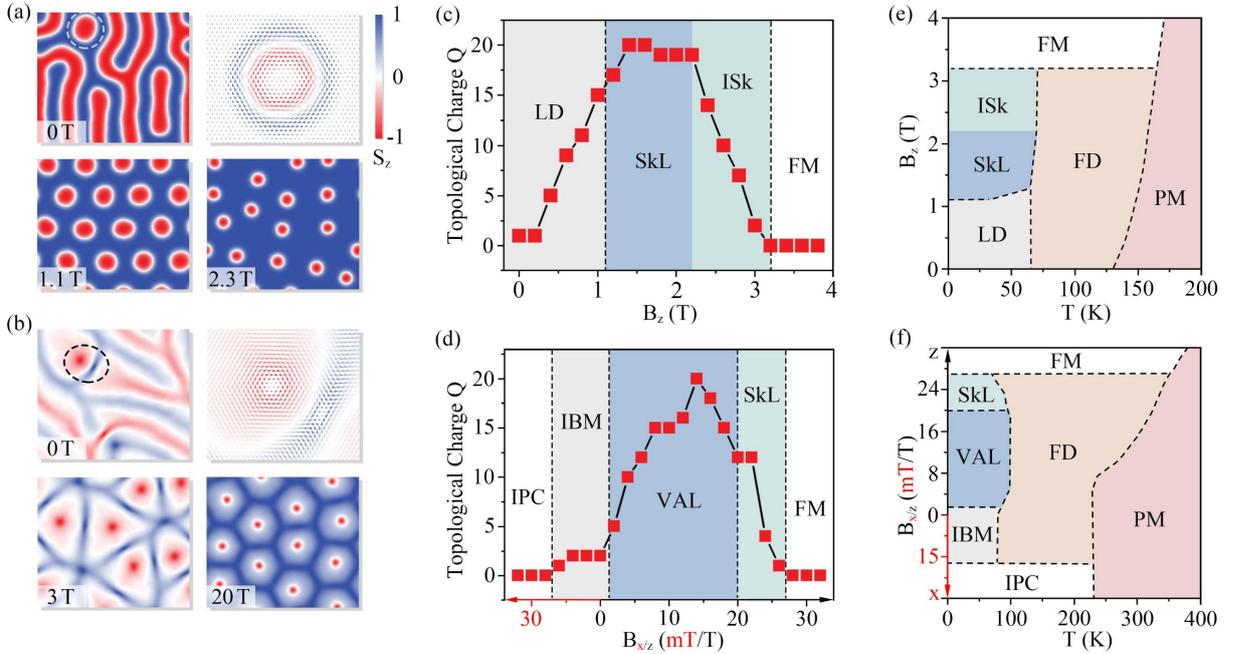

**Figure 3**. Spin textures of monolayer (a) MnInP$_2$Te$_6$ and (b) MnTlP$_2$Te$_6$ without and with applying magnetic field at 0 K. Top-right panels in (a, b) shows the enlarged spin distribution in (a) skyrmion and (b) bimeron. Colors indicate the out-of-plane spin components and arrows denote the in-plane components. Evolutions of topological charge Q of monolayer (c) MnInP$_2$Te$_6$ and (d) MnTlP$_2$Te$_6$ as a function of magnetic field. Phase diagrams of monolayer (e) MnInP$_2$Te$_6$ and (f) MnTlP$_2$Te$_6$ under temperature and magnetic field.

Next, we perform parallel tempering MC simulations to explore the topological spin textures in monolayer MnXP$_2$Te$_6$ based on the first-principles parametrized Hamiltonian of Eq. (1). Topological charge Q is employed to characterize the non-trivial nature of the spin structures, which can be described as $Q = \frac{1}{4\pi}\int \boldsymbol{m} \cdot \left(\frac{\partial \boldsymbol{m}}{\partial x} \times \frac{\partial \boldsymbol{m}}{\partial y}\right) dxdy$ [52]. Here, $\boldsymbol{m}$ is the normalized magnetization vector. As shown in **Figure 3(a)**, the spin texture with an isolated magnetic skyrmion and labyrinth domains is observed in monolayer MnInP$_2$Te$_6$ in the absence of external magnetic field. Such a mixed phase is referred to as labyrinth domain (LD) phase below. Remarkably, from the enlarged magnetic skyrmion shown in the top-right panel of **Figure 3(a)**, we can see that the spins rotate in a plane perpendicular to the radial direction. This confirms that the long-sought Bloch-type magnetic skyrmions are achieved in monolayer MnInP$_2$Te$_6$. The underlying physics is related to the particularly DMI interaction discussed above. While for monolayer MnTlP$_2$Te$_6$, as shown in **Figure 3(b)**, the spin texture with isolated Bloch-type magnetic bimerons (IBM phase) appears in the absence of magnetic field. The difference between them can be attributed to the different magnetic anisotropies.

We then investigate the influence of magnetic field on the topological spin textures of monolayer MnXP$_2$Te$_6$. For monolayer MnInP$_2$Te$_6$, with increasing out-of-plane magnetic field ($B_z$), the labyrinth domains gradually shrink and split into more skyrmions; see **Figure 3(a)**. When $B_z$ reaches at 1.1 T, the labyrinth domains completely disappear, forming skyrmion lattice (SkL) phase. Such intriguing SkL phase can be maintained within the magnetic field range of 1.1-2.2 T. When further increasing $B_z$ to 2.2-3.2 T, the ordered array of skyrmions is destroyed, forming isolated skyrmion (ISk) phase. Concerning the density of skyrmions, as shown in **Figure 3(c)**, it increases as the $B_z$ increases from 0 T to 1.4 T and remains mostly steady under 1.4-2.2 T. Upon increasing $B_z$ larger than 2.2 T, the density of magnetic skyrmions decreases significantly, and drops to zero at 3.2 T, which corresponds to the trivial FM phase. The physics underlying this evolution arises from compensation effect of $B_z$ that enhances spin collinear arrangement.

Unlike monolayer MnInP$_2$Te$_6$, due to the large in-plane magnetization, the topological spin textures of monolayer MnTlP$_2$Te$_6$ are not so sensitive to $B_z$. As shown in **Figure 3(d)**, the IBM phase can be preserved within the range of 0-1.5 T. By further increasing $B_z$ larger than 1.5 T, the vortex-antivortex loops (VAL) appear; see **Figure 3(b)**. With increasing $B_z$ to 0-14 T, the topological charge Q rises gradually, corresponding to an increase in the density of vortices. This phenomenon is attributed to the fact that $B_z$ weakens the energetic advantage of in-plane magnetization contributed by magnetic anisotropy. In VAL region, with increasing $B_z$ from 1.5 to 20 T, the blue vortices and antivortices gradually expand into FM background, and eventually the red vortices evolve into Bloch-type skyrmions at 20 T, forming the SkL phase. As shown in **Figure 3(d)**, under $B_z$ of 14-27 T, the topological charge Q decreases gradually, and reduce to zero at 27 T, corresponding to the transformation into trivial FM phase. In contrast the cases of $B_z$, topological spin textures in monolayer MnTlP$_2$Te$_6$ are more sensitive to the in-plane magnetic field ($B_x$). We can see from **Figure 3(d)** that the IBM phase can be stable only within the $B_x$ range of 0-18 mT. Upon increasing $B_x$ larger than 18 mT, the bimerons are completely magnetized into trivial spin pattern superimposed by in-plane cycloidal structure (IPC phase) and small out-of-plane waved spin pattern [25], see **Figure S7**.

To get further insight into the thermal stability of these topological spin structures, we establish the T-B magnetic phase diagrams, which are displayed in **Figures 3(e,f)**. In addition to various magnetic phases mentioned above, two extra phases emerge under temperature, i.e., fluctuation-disorder (FD) phase (wherein the topological stability of spin configuration is damaged by thermal fluctuations) and paramagnetic (PM) phase. Note that in FD and PM phases, the topological spin textures are destabilized by thermal fluctuations, and thus topological charge Q oscillates with a large amplitude. Here, $||Q| - 1| > 0.01$ per skyrmion (or pair of vortices) is used as the judgment criterion for the transformation into FD phase. In addition, the Curie temperatures under different external magnetic fields are estimated to judge the critical temperature of phase transition into PM phase. Importantly, as illustrated in **Figures 3(e,f)**, the Bloch-type topological magnetisms of monolayer MnInP$_2$Te$_6$ and MnTlP$_2$Te$_6$ realized under different magnetic fields [e.g., LD, SkL, ISk, IBM and VAL phases] can sustain the temperature up to ~70 and 80 K, respectively. With further increasing temperature, topological magnetisms are transformed into FD and PM phases successively. Exceptionally, for monolayer MnInP$_2$Te$_6$ under $B_z$ of 1.1-1.3 T, with increasing temperature, the SkL phase is transformed into LD phase first, then FD phase and finally PM phase; see **Figure 3(e)**. In general, these Bloch-type topological magnetisms can exist in a relative wide window of the T-B phase diagrams.

At last, we study the strain effect on the magnetic parameters as well as topological magnetisms of monolayer MnXP$_2$Te$_6$. As displayed in **Figures 4(a,b)**, the magnetic parameters of monolayer MnXP$_2$Te$_6$ show significant variation as a function of strain. In this regard, the topological spin textures of monolayer MnXP$_2$Te$_6$ are expected to vary significantly with strain. **Figure 4(c)** displays the spin textures of monolayer MnInP$_2$Te$_6$ under different strain and $B_z$. Under tensile strain, the LD phase remains stable, but no magnetic skyrmion is observed under zero magnetic field. And with the increase of tensile strain, the density of the labyrinth domains increases, and the $B_z$ required for realizing magnetic skyrmions increases gradually. Note that $J$ and $B_z$ are prone to force the parallel arrangement of magnetic moments, while $d_\parallel$ induces the magnetic moments in a noncollinear arrangement. As shown in **Figure 4(a)**, increasing tensile strain leads to the increase of $d_\parallel / J$, indicating that DMI plays a more dominated role under tensile strain, which is profit for the emergence of more labyrinth domains.

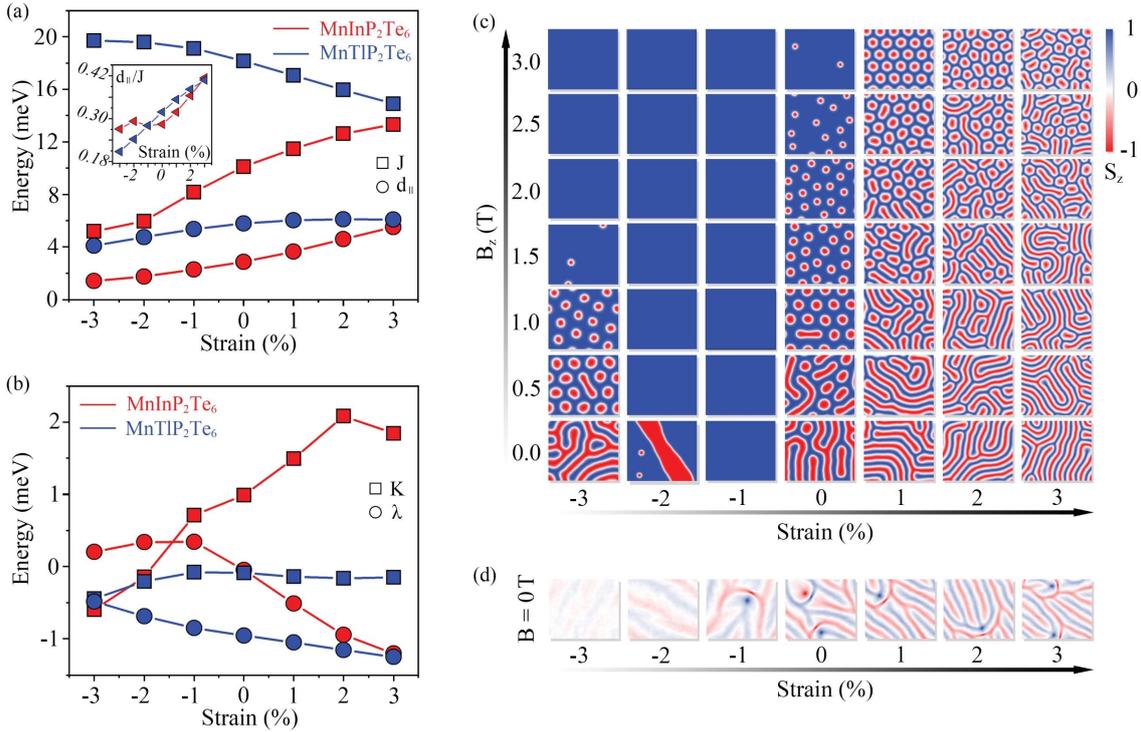

**Figure 4** (a) Magnetic parameters $J$ and $d_\parallel$ of monolayer MnXP$_2$Te$_6$ as a function of strain. Inset in (a) shows $d_\parallel / J$ of monolayer MnXP$_2$Te$_6$ as a function of strain. (b) Magnetic parameters $K$ and $\lambda$ of monolayer MnXP$_2$Te$_6$ as a function of strain. Spin textures of monolayer (c) MnInP$_2$Te$_6$ and (d) MnTlP$_2$Te$_6$ as a function of strain. Colors in (c,d) indicate the out-of-plane spin components.

For the case under compressive strain, the change of $d_\parallel / J$ is relatively slight, and hence the magnetic anisotropy determined by $K$ and $\lambda$ dominates the evolution of spin patterns with strain. With increasing compressive strain to -1%, $\lambda$ increases significantly, while $K$ decreases relatively moderate, resulting in a significant increase of out-of-plane magnetic anisotropy; see **Figures 4(b)** and **S8**. As magnetic anisotropy tends to force the magnetic moments to be arranged parallelly, this makes the labyrinth domains and magnetic skyrmions transformed into the FM phase in the absence of magnetic field [**Figure 4(c)**]. When further increasing the compressive strain, $K$ and $\lambda$ decrease simultaneously, resulting in the decrease of out-of-plane magnetic anisotropy. From **Figure 4(c)**, we can see that under -2% strain, the ISk phase containing isolated magnetic skymions and a strip domain wall appears at 0 T, and can be transformed into FM phase by applying $B_z$. It is interesting to point that skyrmion motion can be enhanced by the domain wall, and the accompanying skyrmion Hall effect can thus be suppressed [53]. When further increasing compressive strain to -3%, the LD phase becomes stable again, which results from the continuous decrease of out-of-plane magnetic anisotropy.

For monolayer MnTlP$_2$Te$_6$, with increasing strain, $d_\parallel$ increases slightly and $K$ is almost unchanged. Unlike $d_\parallel$ and $K$, $J$ and $\lambda$ decrease significantly, leading to a monotonic increase of both $d_\parallel / J$ and in-plane magnetic anisotropy; see **Figures 4(a,b)** and **S8**. **Figure 4(d)** shows the evaluation of spin textures of monolayer MnTlP$_2$Te$_6$ under various strain. It can be seen that the IBM phase of monolayer MnTlP$_2$Te$_6$ in the absence of magnetic field can be maintained within the strain range of -1% – 3%, while the cases under -3% – -1% strain are transformed into trivial spin patterns. Accordingly, by applying moderate strain, the topological magnetism is controllable in monolayer MnXP$_2$Te$_6$.

**Conclusion**

To summarize, using first-principles calculations and MC simulations, we report the identification of Bloch-type magnetic skyrmions in two-dimensional lattice of MnInP$_2$Te$_6$. Meanwhile, Bloch-type magnetic bimerons are also observed in monolayer MnTlP$_2$Te$_6$. The underlying physics of such Bloch-type spin configurations are related to the D$_3$ symmetry. The phase diagrams of these topological magnetisms under magnetic field, temperature and strain are mapped out. Our findings significantly promote the research on topological magnetism in two-dimensional lattice.

**Supporting Information**

The Supporting Information is available free of charge at []

**Competing Interests**

The authors declare no competing interests.

**Acknowledgement**

This work is supported by the National Natural Science Foundation of China (Nos. 12274261 and 12074217), Shandong Provincial Science Foundation for Excellent Young Scholars (No. ZR2020YQ04), Shandong Provincial Natural Science Foundation (Nos. ZR2019QA011), Shandong Provincial Key Research and Development Program (Major Scientific and Technological Innovation Project) (No. 2019JZZY010302), Shandong Provincial QingChuang Technology Support Plan (No. 2021KJ002), and Qilu Young Scholar Program of Shandong University.